\documentclass[12pt]{article} 

\usepackage{amssymb} 
\usepackage{amsmath}
\usepackage{amsfonts}

\newcommand{\R}{\mathbb{R}}
\newcommand{\C}{\mathbb{C}}

\newcommand{\be}{\begin{equation}}
\newcommand{\bea}{\begin{eqnarray}}
\newcommand{\eea}{\end{eqnarray}}
\newcommand{\nn}{\nonumber}

\newcommand{\ed}{\end{document}}

\begin{document}

\title{On a Factorization of Symmetric Matrices and Antilinear Symmetries}
\author{Ali Mostafazadeh\thanks{E-mail address: amostafazadeh@ku.edu.tr}\\ \\
Department of Mathematics, Ko\c{c} University,\\
Rumelifeneri Yolu, 80910 Sariyer, Istanbul, Turkey}
\date{ }
\maketitle

\begin{abstract}
We present a simple proof of the factorization of (complex) symmetric matrices into a product of a square matrix and its transpose, and discuss its application in establishing a uniqueness property of certain antilinear operators.
\end{abstract}

\baselineskip=24pt

\section{Introduction}
One of the interesting results of linear algebra is that every square matrix may be factored into the product of two symmetric matrices \cite{bosch,prasolov}. The factorization of symmetric matrices into a product of a square matrix and its transpose is however less known. In fact, there seems to be no mention of this factorization in modern texts on linear algebra. The purpose of this note is to
present a simple derivation of this particular factorization of symmetric matrices and to discuss its application in establishing a uniqueness property of certain antilinear operators.

\section{Notation and Definitions}
In this note we shall express the complex-conjugate, the transpose, and the conjugate-transpose (adjoint) of a matrix (an operator)  $m$ by $\bar m$, $m^T$, and $m^*$, respectively, and identify the elements of $\C^\ell$ by columns of $\ell$ complex numbers. Then the Euclidean inner-product of  two vectors $\vec w_1$ and $\vec w_2$ takes the form $\vec w_2^*\vec w_1$. 
\begin{itemize}
\item[] {\bf Definition~1:} Let $S$ be a set, then the Kronecker delta function $\delta:S^2\to\{0,1\}$ is defined by
	\[\forall a,b\in S,~~~~~~~\delta(a,b)=\delta_{ab}:=\left\{\begin{array}{ccc}
	1 & {\rm for} & a=b\\
	0 & {\rm for} & a\neq b.\end{array}\right.\]
\item[] {\bf Definition~2:} A function ${\cal X}:{\cal H}\to{\cal H}$ acting in a complex vector space ${\cal H}$ is said to be an antilinear operator if for all $x,y\in\C$ and $\phi,\psi\in{\cal H}$, ${\cal X}(x\phi+y\psi)=\bar x{\cal X}\phi+\bar y{\cal X}\psi$, where ${\cal X}\phi$ means ${\cal X}(\phi)$.
\item[] {\bf Definition~3:} An antilinear operator ${\cal X}:{\cal H}\to{\cal H}$ acting in a complex inner-product space ${\cal H}$ with inner-product $(~,~)$ is said to be symmetric or Hermitian if for 
all $\phi,\psi\in {\cal H}$, $({\cal X}\phi,\psi)=({\cal X}\psi,\phi)$.
\item[] {\bf Definition~4:} A linear operator $H:{\cal H}\to{\cal H}$ acting in a complex inner-product space ${\cal H}$ is said to have a symmetry generated by a function $X:{\cal H}\to{\cal H}$
or simply a $X$-symmetry if $H$ and $X$ commute, i.e., $[H,X]:=HX-XH=0$, where $0$ stands for
the zero operator. A symmetry generated by an antilinear operator is called an antilinear symmetry.
\item[] {\bf Definition~5:} Let  $H:{\cal H}\to{\cal H}$ be a linear operator acting in a complex inner-product space ${\cal H}$ and $G:{\cal H}\to{\cal H}$ be an Hermitian invertible linear or antilinear operator. Then $H$ is said to be $G$-Hermitian \cite{azizov} or $G$-pseudo-Hermitian \cite{pt-1} if $H^*=GHG^{-1}$. 
\end{itemize}

\section{Motivation: Consequences of antilinear symmetries} 
Consider a diagonalizable linear operator $H:{\cal H}\to{\cal H}$ acting in a finite-dimensional complex inner-product space ${\cal H}$ with inner-product $(~,~)$. Let $n$ label the eigenvalues $E_n$ of $H$, $\mu_n$ be the multiplicity of $E_n$, and $\psi_{n,a}$ be the eigenvectors corresponding to the eigenvalue $E_n$ where $a\in\{1,2,\cdots,\mu_n\}$ is the degeneracy label. Then it is well-known \cite{f} that the adjoint $H^*$ of $H$ is diagonalizable; the eigenvalues $\tilde E_n$ of $H^*$ are complex conjugate of those of $H$, i.e., $\tilde E_n=\bar E_n$; the multiplicity of $\tilde E_n$ is equal to $\mu_n$; and one can choose the eigenvectors $\phi_{n,a}$ of $H^*$ in such a way that for all spectral labels $m,n$ and degeneracy labels $a,b$, $( \phi_{n,a},\psi_{n,a})=
\delta_{n,m}\delta_{a,b}$. Clearly, both sets of eigenvectors $\psi_{n,a}$ of $H$ and eigenvectors $\phi_{a,n}$ of $H^*$ form bases of ${\cal H}$;
$\{\psi_{n,a},\phi_{n,a}\}$ is a complete biorthonormal system. Recently \cite{pt-3}, we have shown that if the eigenvalues of $H$ are real, then $H$ has an antilinear symmetry. More generally, we proved the following theorem.
	\begin{itemize}
	\item[]{\bf Theorem~1:} The presence of an antilinear symmetry of $H$ is a necessary and 	sufficient condition for the eigenvalues of $H$ to either be real or come in complex conjugate 	pairs.
	\end{itemize}	
The proof of Theorem~1 uses the following lemma.
	\begin{itemize}
	\item[]{\bf Lemma~1:} Every diagonalizable linear operator $H:{\cal H}\to {\cal H}$ acting in a 
	finite-dimensional complex inner-product space ${\cal H}$ is ${\cal T}$-Hermitian,
		\be
		H^*={\cal T}H{\cal T}^{-1},
		\label{1}
		\end{equation}
	for some Hermitian, invertible, antilinear operator ${\cal T}:{\cal H}\to {\cal H}$.
	\end{itemize}
It turns out \cite{pt-3} that any such ${\cal T}$ may be expressed in terms of the eigenvectors 
$\phi_{n,a}$ of $H^*$ according to
	\be
	\forall \zeta\in{\cal H},~~~~~~~~~~
	{\cal T}\zeta=\sum_n\sum_{a,b=1}^{\mu_n}c_{ba}^{(n)}
	(\phi_{n,a},\zeta)\phi_{n,b},
	\label{T}
	\end{equation}
where $c^{(n)}_{ab}$ are the entries of symmetric invertible $\mu_n\times\mu_n$ matrices $c^{(n)}$. 

Note that Theorem~1 and Lemma~1 have infinite-dimensional generalizations for linear operators $H$ admitting a complete biorthonormal system of eigenvectors \cite{pt-3}.

Next, consider a general basis transformation,
	\be
	\phi_{n,a}\to\phi'_{n,a}:=\sum_{b=1}^{\mu_n}v^{(n)}_{ba}\phi_{n,b}
	\label{bt}
	\end{equation}
where $v^{(n)}_{ab}$ are the entries of an invertible $\mu_n\times\mu_n$ matrix $v^{(n)}$.
In terms of the transformed basis vectors ${\cal T}$ has the form: $\forall \zeta\in{\cal H}$,
${\cal T}\zeta=\sum_n\sum_{a,b=1}^{\mu_n}c_{ba}^{'(n)}(\phi'_{n,a},\zeta)\phi'_{n,b}$,
where $c^{'(n)}$ are related to $c^{(n)}$ according to $c^{(n)}=v^{(n)}c^{'(n)}v^{(n)T}.$
This equation indicates that the issue of the uniqueness of ${\cal T}$ for a given $H$ is related to
whether one can find for each $c^{(n)}$ an invertible matrix $v^{(n)}$ such that
$c^{(n)}=v^{(n)}v^{(n)T}$. In the remainder of this note we shall give a proof of the fact that this is indeed possible, and one can transform to a basis in which ${\cal T}$
has the (canonical) form (\ref{T}) with $c^{(n)}_{ab}=\delta_{ab}$ for all $n$.

\section{Factorization of Symmetric Matrices}
	\begin{itemize}
	\item[] {\bf Theorem~2:} A square matrix $c$ is symmetric if and only if it can be written as 	$c=vv^{T}$ for some square matrix $v$.
	\item[] {\bf Proof:} If $c=vv^{T}$ then clearly $c$ is symmetric. To prove the converse we 	use induction on the dimension $n$ of the matrix $c$. For $n=1$, $c=vv^{T}=v^2$ is trivially 
	satisfied by letting $v:=\sqrt{c}$. By induction hypothesis we assume that for all  	$k\in\{2,\cdots,n\}$, every $k\times k$ symmetric matrix $c$ can be written in the form
	$c=vv^{T}$ for some $k\times k$ matrix $v$. 
	Now let $C$ be an $(n+1)\times(n+1)$ symmetric matrix. Then $C$ has at least one 
	eigenvector \cite{gelfand} i.e., there are $\vec e\in\C^{n+1}-\{\vec 0\}$ and $\lambda\in\C$ such 
	that 
		\be
		C \vec e=\lambda \vec e.
		\label{eg-va}
		\end{equation}
	Now let ${\cal V}:=\{\vec w\in\C^{n+1}| \vec w^*\overline{\vec e}=0\}$ be the orthogonal 
	complement	of $\overline{\vec e}$. Clearly ${\cal V}$ is an $n$-dimensional vector subspace of 
	$\C^{n+1}$. Next, consider the following two possibilities.
		\begin{itemize}
		\item[(i)]  $\vec e\notin{\cal V}$. In this case, choose a basis 
	$\{\vec e_1,\vec e_n,\cdots,\vec e_n\}$ of	${\cal V}$ and let $\vec e_{n+1}:=\vec e$. Then 
	$\{\vec e_1,\vec e_2,\cdots,\vec e_n,\vec e_{n+1}\}$ is a 
	basis 	of $\C^{n+1}$ and the matrix $A:=(\vec e_1,\vec e_n,\cdots,\vec e_n,\vec e_{n+1})$ is 
	invertible. Note that for all $\ell\in\{1,2,\cdots,n\}$, $\vec e_\ell\in V$, and 
	$\vec e_\ell^{\: T}\vec e_{n+1}=0$. This in turn implies that the matrix $A^T C A$ which is
 	symmetric has the block form
		\be
		A^TCA=\left(\begin{array}{cc}
				\tilde c & \vec 0\\
				\vec 0^{\: T} & \lambda^2\end{array}\right),
		\label{ACA}
		\end{equation}
	where $\tilde c$ is a symmetric $n\times n$ matrix. By induction hypothesis there is an
	$n\times n$ matrix $\tilde v$ such that $\tilde c=\tilde v\tilde v^T$.  Now let $B$ be the
	$(n+1)\times(n+1)$ matrix 
		\be
		B:=\left(\begin{array}{cc}
				\tilde v^T& \vec 0\\
				\vec 0^{\: T} & \lambda \end{array}\right),
		\label{B}
		\end{equation}
	and $V:=(BA^{-1})^T$. Then in view of (\ref{ACA}) and (\ref{B}), $B^TB=A^TCA$ and
		\[ VV^T=(BA^{-1})^TBA^{-1}=A^{-1T}B^TB A^{-1}=C.\]
	This completes the proof for case (i).
	\item[(ii)] $\vec e\in{\cal V}$, i.e., $\vec e^{\: T}\vec e=0$. In this case, let 
	${\cal V}':=\{\vec w\in{\cal V}|	\vec w^*\vec e=0\}$ be the orthogonal complement of 
	$\vec e$ in ${\cal V}$, 
	$\{\vec e_1,\vec e_2,\cdots,\vec e_{n-1}\}$ be a basis of ${\cal V}'$, $\vec e_n=
	\overline{\vec e}$,
	and $\vec e_{n+1}=\vec e$. Then $\{\vec e_1,\vec e_2,\cdots,\vec e_n,\vec e_{n+1}\}$ is a 
	basis of
	$\C^{n+1}$ and the matrix $A':=(\vec e_1,\vec e_n,\cdots,\vec e_n,\vec e_{n+1})$ is 
	invertible. Note that for all $\ell\in\{1,2,\cdots,n-1\}$, $\vec e_\ell\in V$, and 
	$\vec e_\ell^{\: T}\vec e_{n+1}=0$. Furthermore, $\vec e_{n+1}^{\: T}\vec e_{n+1}=
	\vec e^{\: T}\vec e=0$ and $\alpha:=\vec e_{n}^{\: T}\vec e_{n+1}=
	\vec e^{*}\vec e\in\R^+$. In view of these relations and (\ref{eg-va}), 
		\be
		C':=A^{'T}CA'=\left(\begin{array}{cccccc}
		\tilde c'_{1,1} & \tilde c'_{1,2} &\cdots&\tilde c'_{1,n-1}&\tilde c'_{1,n}&0\\
		\tilde c'_{2,1} & \tilde c'_{2,2} &\cdots&\tilde c'_{2,n-1}&\tilde c'_{2,n}&0\\
		\vdots & \vdots &\ddots&\vdots&\vdots& 0\\
		\tilde c'_{n-1,1} & \tilde c'_{n-1,2} &\cdots&\tilde c'_{n-1,n-1}&\tilde c'_{n-1,n}&0\\
		\tilde c'_{n,1} & \tilde c'_{n,2} &\cdots&\tilde c'_{n,n-1}&\tilde 		c'_{n,n}&\lambda\alpha\\
		0&0&\cdots&0&\lambda\alpha&0\end{array}\right),
		\label{ACA-n}
		\end{equation}
	where $\tilde c'_{i,j}:=\vec e_i^{\: T}C\vec e_j$ are the entries of a symmetric $n\times n$ matrix 
	$\tilde c'$. Now if $\lambda=0$, $C'$ is block-diagonal and the argument given in case (i) leads 
	to a proof of the theorem.
	This leaves the case $\lambda\neq 0$. In this case, let $A:=A'D$ where
	$D$ is the $(n+1)\times(n+1)$ matrix 
		\be
		D=\left(\begin{array}{ccccccc}
			1&0&0&\cdots&0&0&x_1\\
			0&1&0&\cdots&0&0&x_2\\
			\vdots&\vdots&\vdots&\ddots&\vdots& \vdots&\vdots\\
			0&0&0&\cdots&1&0&x_{n-1}\\
			0&0&0&\cdots&0&1&x_{n}\\
			y_1&y_2&y_3&\cdots&y_{n-1}&y_{n}&0
			\end{array}\right),
		\label{D}
		\end{equation}
	with $x_1,x_2,\cdots x_{n-1}$ being arbitrary complex numbers, 
		\be
		x_n:=-(\lambda\alpha)^{-1},
		\label{xn}
		\end{equation}
	and for all $i\in\{1,2,\cdots,n\}$
		\be
		 y_i:=\sum_{j=1}^n \tilde c'_{i,j} x_j.
		\label{yi}
		\end{equation}
	Then a simple computation shows that
		\[A^TCA=D^TC'D=\left(\begin{array}{cc}
		\tilde c_{ij}&\vec 0\\
		\vec 0^{\:T} &\lambda^{'2}\end{array}\right),\]
	where $\tilde c_{i,j}$ are the entries of a symmetric $n\times n$ matrix $\tilde c$ and 	$\lambda'\in\C$. Therefore, we can use the argument given in case (i) to show the
	existence of an $(n+1)\times(n+1)$ matrix $B$ satisfying 
		\be
		A^TCA=B^TB.
		\label{DCD}
		\end{equation}
	The proof of the theorem will be complete if we show that the matrix $A=A'D$ is invertible. 	Because $A'$ is invertible, it suffices to show the existence of $x_1,x_2,\cdots x_{n-1}$ for 	which $\det D\neq 0$. We can use the properties of the determinant and Equations~(\ref{xn}) 
	and (\ref{yi}) to compute
		\bea
		\det D&=&-\sum_{i=1}^{n}x_iy_i\nn\\
			&=&-\sum_{i=1}^{n-1}\tilde c'_{i,i} x_i^2-2\sum_{i<j=1}^{n-1}
			\tilde c'_{i,j}x_ix_j+2(\lambda\alpha)^{-1}\sum_{i=1}^{n-1}
			\tilde c'_{n,i}x_i-(\lambda\alpha)^{-2}\tilde c'_{n,n}.
		\label{det}
		\eea
	Suppose that for all values of $x_1,x_2,\cdots x_{n-1}$, $\det D=0$. This implies that
	$\tilde c'=0$  in which case $C'$ will have the form
		\[ C'=	\left(\begin{array}{cccccc}
			0&0&\cdots &0&0&0\\
			0&0&\cdots&0&0&0\\
			\vdots&\vdots&\ddots&\vdots&\vdots&\vdots\\
			0&0&\cdots&0&0&0\\
			0&0&\cdots&0&0&\lambda\alpha\\
			0&0&\cdots&0&\lambda\alpha&0
			\end{array}\right).\]
	By induction hypothesis we have a $2\times 2$ matrix $m$ satisfying
		\[ 	\left(\begin{array}{cc}
			0&\lambda\alpha\\
			\lambda\alpha&0
			\end{array}\right)=m^Tm.\]
	Therefore, setting
		\[ B:=	\left(\begin{array}{cccccc}
			0&0&\cdots &0&0&0\\
			0&0&\cdots&0&0&0\\
			\vdots&\vdots&\ddots&\vdots&\vdots&\vdots\\
			0&0&\cdots&0&0&0\\
			0&0&\cdots&0&m_{1,2}&m_{1,2}\\
			0&0&\cdots&0&m_{2,1}&m_{2,2}
			\end{array}\right)\]
	and $V:=(BA^{'-1})^T$, we have $C'=B^TB$ and $C=VV^T$. If
	$\tilde c'\neq 0$ then there are values of $x_1,x_2,\cdots x_{n-1}$ for which $D$
	is a nonsingular matrix and $A=A'D$ is invertible. Hence we can set $V:=(BA^{-1})^T$
	and use (\ref{DCD}) to show that $C=VV^T$.~~$\square$
	\end{itemize}
	\end{itemize}

\section{Concluding remarks}

\begin{enumerate}
\item The factorization $c=vv^T$ established in Theorem~2 is invariant under the transformation
$v\to v'=vo$ where $o$ is an arbitrary (complex) orthogonal matrix. In particular, one may choose
$o$ so that the factorizing matrix $v'$ has a simple form.
\item In view of the discussion of Section~3, one has the following consequence of Theorem~2.
	\begin{itemize}
	\item[] {\bf Corollary~1:} Up to basis transformations~(\ref{bt}), there is a unique antilinear 	operator ${\cal T}$ satisfying Equation~(\ref{1}), namely
		\be
		\forall \zeta\in{\cal H},~~~~~~~~~~
		{\cal T}\zeta=\sum_n\sum_{a=1}^{\mu_n}(\phi_{n,a},\zeta)\phi_{n,a},
		\label{T2}
		\end{equation}
	\end{itemize}
\item For a self-adjoint linear operator $H$, one can set $\phi_{n,a}=\psi_{n,a}$ and use the completeness of the eigenvectors $\psi_{n,a}$ and (\ref{T2}) to deduce ${\cal T}^2=I$, where $I$ is the identity operator. Furthermore, noting that in this case (\ref{1}) is equivalent to ${\cal T}$-symmetry of $H$, one can prove the following.
	\begin{itemize}
	\item[] {\bf Corollary~2:} Every self-adjoint linear operator $H$ has an antilinear symmetry 
	generated by a Hermitian, invertible, antilinear operator ${\cal T}$ satisfying ${\cal T}^2=I$.
	\end{itemize}
\end{enumerate}

\section*{Acknowledgment}
This project was supported by the Young Researcher Award Program (GEBIP) of the Turkish Academy of Sciences.

\ed
\begin{thebibliography}{9}
\bibitem{azizov} T.~Ya.~Azizov and I.~S.~Iokhvidov, {\em Linear Operators in Spaces with an Indefinite Metric,}
John Wiley \& Sons, Chichester, 1989.
\bibitem{bosch} A.~J.~Bosch, `The factorization of a square matrix into two symmetric matrices,'
Am.\ Math.\  Monthly {\bf 93}, 462-464 (1986)
\bibitem{f} F.~Chatelin, {\em Eigenvalues of Matrices,} John Wiley \& Sons, Chichester, 1993.
\bibitem{gelfand} I.~M.~Gelfand, {\em Lectures on Linear Algebra,} Dover, New York, 1989.
\bibitem{pt-1} A.~Mostafazadeh, `Pseudo-Hermiticity versus PT-Symmetry I: The necessary condition for the reality of the spectrum of a non-Hermitian Hamiltonian,' J.~Math.\ Phys.\ {\bf 43}, 205-214 (2002).
\bibitem{pt-3} A.~Mostafazadeh, `Pseudo-Hermiticity versus PT-Symmetry III: Equivalence of pseudo-Hermiticity and the presence of antilinear symmetries,' LANL archies preprint: math-ph/0203005.
\bibitem{prasolov} V.~V.~Prasolov, {\em Problems and Theorems in Linear Algebra,} American Mathematical Society, Providence, 1994.
\end{thebibliography}
